\documentstyle[aps,multicol,graphicx,epsfig,amsfonts]{revtex}

\renewcommand{\bar}[1]{\overline{#1}}
\begin{document}   
\newcommand{\z}{\zeta}
\newcommand{\be}{\begin{equation}}
\newcommand{\ee}{\end{equation}}
\parindent = 0pt
\bibliographystyle{prsty}
\tightenlines
\widetext  
\title{ 
Universal eigenvector statistics in a quantum scattering ensemble
}   
\author{B. Mehlig and M. Santer}
\address{Theoretische Quantendynamik, Institut f\"ur Physik,
Albert-Ludwigs Universit\"at Freiburg, Germany}
\date{\today}
\maketitle{ } 
\begin{abstract}   
We calculate eigenvector statistics
in an ensemble of non-Hermitian matrices
describing open quantum systems
[F. Haake {\em et al.}, Z. Phys. B {\bf 88}, 359 (1992)]
in the limit of large matrix size.
We show that ensemble-averaged eigenvector
correlations corresponding to eigenvalues
in the center of the support of
the density of states in the complex
plane are described by an expression
recently derived for Ginibre's ensemble
of random non-Hermitian matrices.
\end{abstract}   
\begin{multicols}{2} 

The statistical properties of
eigenvector overlaps may have an
important bearing on time evolution
and determine the sensitivity
to perturbations of systems governed
by non-Hermitian operators or matrices.
In such systems it is thus important to know the
statistical properties of the (left and right)  
eigenvectors. Despite this fact little is known about eigenvector
correlations in general ensembles
of non-Hermitian random matrices. In Refs. \cite{cha98a,meh99b}, 
eigenvector statistics were calculated for Ginibre's
ensemble of random non-Hermitian matrices
where each matrix element is
an independent, identically distributed
Gaussian complex random variable.
The question arises to which
extent these results are
relevant for other
ensembles of  non-Hermitian
random matrices.

In the following we determine
eigenvector statistics for an ensemble
of non-Hermitian $N\times N$ matrices 
of the form
\begin{equation}
\label{eq:J}
\label{eq:defJ}
J = H + {\rm i}\gamma \sum_{a=1}^M V^a ({{V}^a)}^\dagger\,.
\end{equation}
Here $H$ is a $N\times N$ Hermitian random matrix with
complex, Gaussian distributed matrix elements $H_{kl}$
with zero mean and variance
$ \langle |H_{kl}|^2\rangle = \delta_{kl}\,N^{-1}$\,.
The $V^a$ are vectors with complex random Gaussian
entries $V_k^a$ 
with zero mean and variance
$\langle V_k^a \overline{V}_l^b\rangle = \delta_{kl} \delta_{ab} N^{-1}$.
The eigenvalues, $\lambda_{\alpha}$, of $J$ are 
distributed in the complex plane.
The ensemble (\ref{eq:J}), with $\gamma < 0$,  has been used
to model the statistical properties
of resonances arising in the
case of resonance scattering in open
quantum systems \cite{wei,fyo97b}; the
position of the resonances is modelled
by the real part $\lambda_\alpha'$ of
the eigenvalues of (\ref{eq:defJ}) and
the width by the imaginary part
$\lambda_\alpha''$.
The statistics of eigenvectors for such an ensemble
 was found to be of considerable importance for describing
the properties of random lasing
media, see \cite{lasers}.

The ensemble-averaged density of states
$d(z) = \big\langle N^{-1}\sum_\alpha \delta(z-\lambda_\alpha)\big\rangle$
for the ensemble (\ref{eq:J})
has been worked out using
a number of different techniques,
namely the replica trick \cite{haa92},
the non-linear sigma model approach \cite{fyo97b}
and using the self-consistent Born approximation
\cite{jan97}. 

Very recently, in Ref. \cite{fyo99}, all $n$-point spectral 
correlation functions for a variant of the ensemble
(\ref{eq:defJ}) were determined. It is given by
\be
\label{eq:defJ2}
J = H + {\rm i}\Gamma
\ee
where $H$ is defined as above,
and $\Gamma$ is a fixed, $N\times N$
diagonal matrix with $M$ non-zero
diagonal matrix elements $\gamma$.
For the ensemble (\ref{eq:defJ2}) it was shown,
in particular,
that the spectral two-point function
$R_2(z_1,z_2) = 
 \big\langle N^{-1}\sum_{\alpha\neq\beta} 
 \delta(z_1-\lambda_\alpha)
 \delta(z_2-\lambda_\beta)\big\rangle$
(and all higher correlation functions) are
-- after suitable rescaling and sufficiently far away from the 
boundary of the support of the spectrum --
identical to those derived for Ginibre's ensemble 
(see Eqs. (15.1.31)
and (15.1.37) of Ref. \cite{meh67}). One thus
expects that spectral $n$-point correlations
of the ensemble (\ref{eq:defJ}) are locally
similar to those in Ginibre's ensemble
and thus universal. Furthermore it has 
been argued that under very general circumstances
the fluctuations of the ensembles (\ref{eq:defJ}) and
(\ref{eq:defJ2}) are identical \cite{yan}.

Below, we explore to which extent
the statistical properties of eigen{\em vectors}
in the ensembles (\ref{eq:defJ}) and (\ref{eq:defJ2}) 
are {\em universal}
and the remainder of this paper 
is organized as follows. After defining
the eigenvector correlator to be calculated,
we briefly discuss the method used: the self-consistent
Born approximation. We then derive an expression
for eigenvector correlations and compare
it to results of previous calculations
for Ginibre's ensemble \cite{cha98a,meh99b}.
Finally, we show
results of numerical simulations,
compare them to our analytical results,
and discuss the applicability of
our analytical method.

The eigenvalues of (\ref{eq:defJ})  are non-degenerate with probability one, 
and in this case the left and right eigenvectors, $|L_\alpha\rangle$
and $|R_\alpha\rangle$,
\begin{eqnarray}
\label{eq:eveq}
J\,|R_\alpha\rangle   &=& \lambda_\alpha\,|R_\alpha\rangle\,,\\
\langle L_\alpha|\,J &=& \langle L_\alpha|\,\lambda_\alpha
\nonumber
\end{eqnarray}
form two complete, biorthogonal sets, and can be normalised so that
\be
\label{eq:biorth}
\langle L_\alpha|R_\beta\rangle = \delta_{\alpha\beta}\,.
\ee
We indicate Hermitian conjugates of 
vectors in the usual way, so that, for example, 
$|L_\alpha\rangle$ satisfies $J^{\dagger}\,|L_\alpha\rangle= \bar{\lambda}_\alpha\,|L_\alpha\rangle$.
We investigate the eigenvector correlators \cite{cha98a,meh99b}
\begin{eqnarray}
\label{eq:Odiag}
O(z) &=&  \Big\langle\frac{1}{N}\sum_\alpha O_{\alpha\alpha}
\,\delta(z-\lambda_\alpha)\Big\rangle\,,\\
\label{eq:Ooff}
O(z_1,z_2) &=&\Big\langle \frac{1}{N}\sum_{\alpha\neq\beta} O_{\alpha\beta}
\,\delta(z_1-\lambda_\alpha) \,\delta(z_2-\lambda_\beta)
\Big\rangle
\end{eqnarray}
where $ O_{\alpha\beta} = \langle L_\alpha | L_\beta \rangle\,
                  \langle R_\beta  | R_\alpha\rangle$\,.
These quantities may be extracted from
\be
D(z_1,z_2) 
 =\Big\langle \frac{1}{N}\sum_{\alpha,\beta} O_{\alpha\beta}
\,\delta(z_1-\lambda_\alpha) \,\delta(z_2-\lambda_\beta)
\Big\rangle
\ee
which may be written as
$D(z_1,z_2) = O(z_1)\,\delta(z_1-z_2)+O(z_1,z_2)$.
An expression for the diagonal part $O(z_1)$ for
the ensemble (\ref{eq:defJ}) was derived in \cite{jan99}.
  \begin{figure}
 \narrowtext
 \centerline{\psfig{file=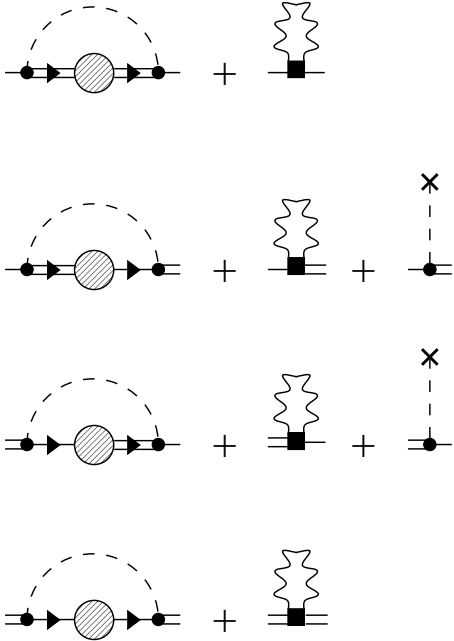,width=6cm}}
 \mbox{}\\[-5mm]
 \caption{\label{fig:dyson} Shows a diagrammatic
 representation of the self-energy $\bbox{\Sigma}$.
 For the diagrammatic rules see
 \protect\cite{meh99b} and  Fig. \protect\ref{fig:diff} below.}
 \end{figure}

\begin{figure}
 \narrowtext
 (a)\hglue2mm
      $\hbox{\protect\raisebox{-6.5mm}
           {\protect\psfig{file=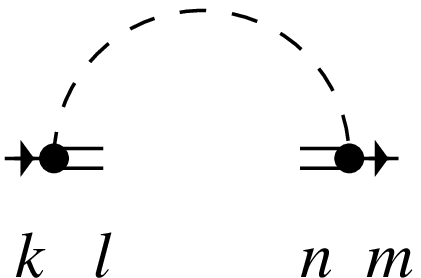,width=2.5cm}}}
 =\langle H_{kl} \overline{H}_{mn}\rangle$
 \mbox{}\\[0.5cm]
 (b)\hglue2mm
      $\displaystyle \hbox{\protect\raisebox{-3.5mm}
           {\protect\psfig{file=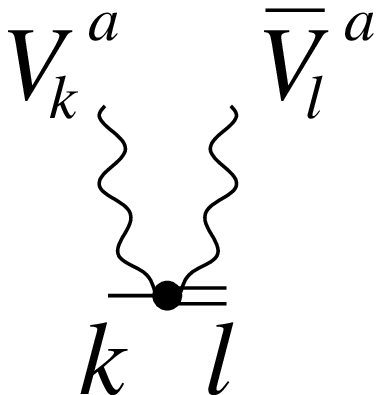,width=1.5cm}}}$
 \hglue-6mm$\displaystyle =i\gamma \sum_{a=1}^M V^a_k \overline{V}^a_l\,,\qquad
      \hbox{\protect\raisebox{-3.5mm}
           {\protect\psfig{file=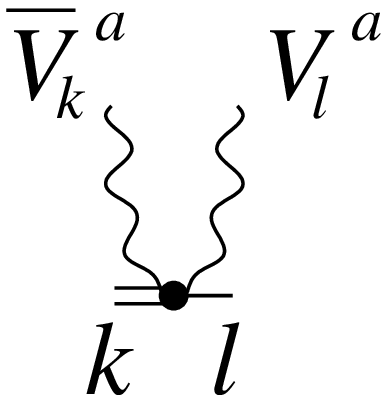,width=1.5cm}}}$
 \hglue-6mm $\displaystyle =-i \gamma \sum_{a=1}^M \overline{V}^a_k V^a_l$\\[5mm]
 (c)\hglue3mm $\hbox{\protect\raisebox{-5.5mm}
           {\protect\psfig{file=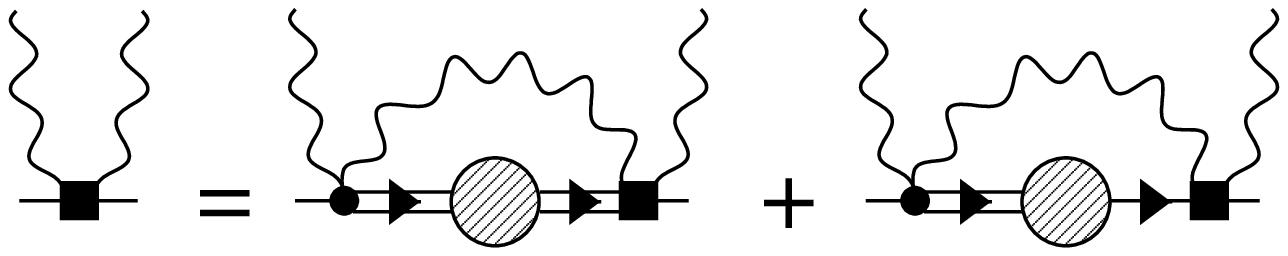,width=7cm}}}$
 \\[5mm]
 \caption{\label{fig:diff} (a) Diagrammatic representation
 of the variance of the matrix $H$ in (\ref{eq:defJ}).
 (b) Representation of the second term of $J$ [Eq. (\ref{eq:defJ})].
 The random variables $V_k^a$ are denoted
 by wavy lines. (c) Shows the self-consistent equations
 determining the vertex
       $\,\,\hbox{\protect\raisebox{-3.5mm}
            {\protect\psfig{file=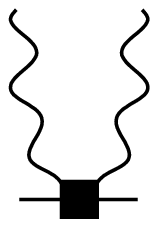,width=.5cm}}}\,.$
There are three more such equations determining the
three remaining vertices occuring in Fig. \ref{fig:dyson}.}
 \end{figure}

{\em Self-consistent Born approximation.} 
We calculate eigenvector correlators 
in terms of averages of products of Green functions,
using an approximation scheme, namely an expansion in powers of $N^{-1}$.
The  method used in Refs. \cite{cha98a,meh99b}, 
yielding exact results for Ginibre's ensemble,
is not readily generalizable.
We use the approach developed in \cite{jan97,cha97,zee97}: 
since the Green functions are non-analytic in the lower (upper)
complex half-plane, a Hermitian $2N\times 2N$ matrix 
$\bbox{H} = \bbox{H}_0 + \bbox{H}_1$ is introduced 
\cite{jan97,cha97}
\begin{equation}
\label{eq:H}
\bbox{H}_0 = 
\left(
\begin{array}{cc}
\eta \,\,& \\  & -\eta
\end{array}
\right )\,,\hspace*{5mm}
\bbox{H}_1 = 
\left(
\begin{array}{cc}
 \,\,& A\\ A^\dagger &
\end{array}
\right )
\end{equation}
with $\eta > 0$, $A = z-J$  and with inverse
\begin{equation}
\label{eq:G}
\bbox{G}
=
\left( 
\begin{array}{cc}
G_{11}& G_{12}\\
G_{21}& G_{22}
\end{array}
\right)\,.
\end{equation}
The resolvents are obtained by taking $\eta \rightarrow 0$.
In this limit, $G_{21} = (z-J)^{-1}$
and $G_{12} = (\overline{z}-J^\dagger)^{-1}$. 
Expanding the Green function $\bbox{G}$ as a power series
in $\bbox{H}_1$, its ensemble average $\langle \bbox{G}\rangle$ 
can be written as
\begin{equation}
\label{eq:sc}
\label{eq:dyson}
\langle\bbox{G}\rangle = \bbox{G}_0 + \bbox{G}_0\bbox{\Sigma}
\langle \bbox{G}\rangle\,,
\end{equation}
where $\bbox{G}_0 = \bbox{H}_0^{-1}$ and $\bbox{\Sigma}$
is a self-energy. 
A graphical representation of the self-energy 
(valid for $M,N$ large and  $M/N \equiv m =\mbox{const.}$)
is
given in Fig. \ref{fig:dyson}. The diagrammatic
rules are analogous to those    described
in \cite{meh99b}. Differences are briefly
explained in Fig. \ref{fig:diff}. 
Eq. (\ref{eq:dyson})
is solved for $\langle \bbox{G}\rangle$ in the limit
 of $\eta\rightarrow 0$.
Expressions for the averaged Green function
are given in \cite{jan97}.

From $\langle G_{21} \rangle$ one obtains the
density of states $d(z)$ in the usual fashion \cite{jan97,cha97,zee97}.
In order to make connection with the results discussed 
in \cite{fyo99}, we specialize
to the limit of small $m$.
In this limit one obtains 
\be 
\label{eq:dos}
d(0,y) \simeq \mbox{const.} + m/(4\pi y^2)
\ee
for $m/(g+1) \leq 2 y \leq m/(g-1)$ and zero
otherwise, compare
Eq. (108) in Ref. \cite{fyo97b}.
Here $2g = (\gamma + 1/\gamma)$
and $z=x+{\rm i}y$ with $x,y$ real.
We will analyze eigenvector correlations in the center $z_0$
of the support of the density of
states, $z_0 = x_0 + {\rm i}y_0$ with $x_0 = 0$
and $y_0 = m/(2g)$ where $d_0 \equiv d(z_0) \simeq g^2/(\pi m)$.
 \begin{figure}
 \centerline{\psfig{file=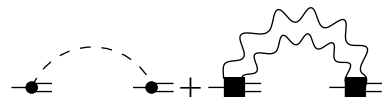,width=5cm}}
 \mbox{}\\[-8mm]
 \caption{\label{fig:bse} Vertex for calculating
 averages of products of Green functions in
 the limit of large $M,N$ with $M=mN$.}
 \end{figure}

{\em Eigenvector correlators}.
We make use of the relation 
\be
\label{eq:deriv}
D(z_1,z_2) = \frac{1}{\pi^2}\frac{\partial }{\partial \overline{z}_1}
             \frac{\partial }{\partial z_2} F(z_1,z_2)
\ee
with
\be
F(z_1,z_2) = \Big\langle N^{-1}\mbox{Tr}\,
\big [(z_1-J)^{-1}\,(\overline{z}_2-J^\dagger)^{-1}\big]
\Big \rangle \,.
\ee
An expression for this average
may be derived as described in \cite{meh99b};
cf. this reference for a diagrammatic representation of $F(z_1,z_2)$.
The only difference between the case of interest
here and the one discussed in Ref. \cite{meh99b} 
is that the vertex must
be replaced by that shown in Fig. \ref{fig:bse}.
The corresponding expression for $F(z_1,z_2)$
is valid in the limit of $M,N$ large, $M=mN$
and for $|z_1 -z_2|^2 > (\pi d(z_+)\, N)^{-1}$ 
with $z_+ = (z_1+z_2)/2$.

Using Eq. (\ref{eq:deriv}) we obtain
a (rather lengthy) expression for $O(z_1,z_2)$. 
To keep the formulae simple, we specialise   further to the case where
$z_1$ and $z_2$ are in the vicinity of $z_0$
and obtain to leading order in $m$
and lowest order in $|\delta z|$
\be
\label{eq:resultO}
O(z_0,z_0+\delta z) \simeq -\left(\frac{m}{g^2} -\frac{m^2}{2g^4}\right)
\frac{1}{\pi^2}\frac{1}{|\delta z|^4}\,.
\ee
Corrections breaking rotational invariance are
of higher order. 
 Comparison with Eq. (8) of Ref. \cite{cha98a}
 shows that locally, near the center of
 the support of the density of states,
 the eigenvector correlations for
 the ensemble (\ref{eq:defJ}) are the
 same as those in Ginibre's ensemble,
 apart from an additional factor of $m/g^2-m^2/(2g^4)$
which measures the strength 
of the non-Hermiticity in Eq. (\ref{eq:defJ}).
According to Eq. (\ref{eq:resultO}), eigenvector
correlations are strongest for $\gamma\simeq 1$,
and vanish in the limit of $\gamma\rightarrow 0$
and $\gamma\rightarrow\infty$ which
correspond to symmetric and complex symmetric
$J$, respectively.

\begin{figure}
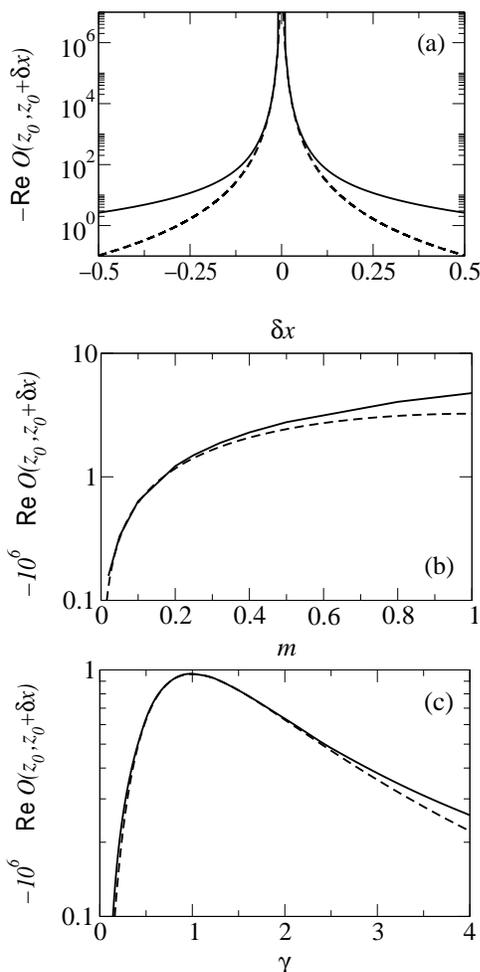

 \parbox{8cm}{%
 \centerline{\hspace*{1mm}%
 \includegraphics[clip,width=0.35\textwidth]{asympt.eps}}

 \centerline{\includegraphics[clip,width=0.35\textwidth]{m.eps}}

 \centerline{\includegraphics[clip,width=0.35\textwidth]{g.eps}}}
\caption{\label{fig:comp} Shows ${\sf Re} O(z_0,z_0+\delta x)$
 as obtained within the self-consistent Born approximation
(solid line)
 compared with the asymptotic expression (\ref{eq:resultO})
 (dashed line)  (a)
 as a function of $\delta x$ for $m=0.1$ and $\gamma=0.5$, (b)
 as a function of $m$ for $\delta x = 0.01$ and $\gamma=0.5$,
(c) as a function of $\gamma$ for $m=0.1$ and $\delta x=0.01$.}
\end{figure}

Given the expression (\ref{eq:resultO}),
we may estimate $O(z_0)$ up to a constant
of order unity, in the way described in
\cite{meh99b}. We obtain, to lowest order
in $m$,
\be
O(z_0) \simeq N \frac{m}{g^2}
\ee
which is consistent with the result
derived in \cite{jan99}.
In Fig.~\ref{fig:comp} we compare
the expression  (\ref{eq:resultO})
to the full result for $O(z_1,z_2)$ -- 
as obtained within the self-consistent
Born approximation -- and
observe very good agreement for $|\delta z|$ 
not too large.

Eq. (\ref{eq:resultO}) is valid
to lowest order in $|\delta z|^{-1}$ and provided
$|\delta z|^2 > (\pi d_0 N)^{-1}$.
The behaviour of $O(z_0,z_0+\delta z)$
for smaller values of $\delta z$ may
be understood as follows.
Assuming
\be
O(z_0,z_0+\delta z)
\simeq \langle O_{\alpha\beta}\rangle
\,R_2(z_0,z_0+\delta z)
\ee
for $|\delta z| \simeq |\lambda_\alpha-\lambda_\beta|$
very small, one may estimate the two factors
on the r.h.s. of this equation separately.
First, if two eigenvalues
$\lambda_\alpha$
and $\lambda_\beta$ of $J$ are very close to each other,
one may argue that
the corresponding overlap matrix element $O_{\alpha\beta}$
scales as $|\lambda_\alpha-\lambda_\beta|^{-2}$.
This is seen by considering
a $2\times 2$ matrix $J$ with arbitrary
complex matrix elements $J_{kl}$. Denoting
its  right eigenvectors by
$|R_\alpha\rangle = (1,\overline{\varrho}_\alpha)^\dagger$,
the corresponding left eigenvectors, assuming
$\lambda_1 \neq \lambda_2$ and subject
to the condition (\ref{eq:biorth}) of biorthogonality,  are
given by
by $\langle L_1 | = (-\varrho_2,1)/(\varrho_1-\varrho_2)$
and $\langle L_2 | = (-\varrho_1,1)/(\varrho_2-\varrho_1)$. Thus
\begin{eqnarray}
O_{12} &\equiv &
\langle L_1 | L_2\rangle 
\langle R_2| R_1\rangle \propto -|\varrho_1-\varrho_2|^{-2}
\\
&\propto& -|\lambda_1-\lambda_2|^{-2}\,.
\nonumber
\end{eqnarray}
For $\lambda_\alpha$ very close to $\lambda_\beta$
[namely on scales smaller than the mean level spacing $(\pi d_0 N)^{-1/2}$]
this behaviour pertains for arbitrary
values of $N$.
Second, the spectral two-point function scales as
$R_2(z_0,z_0+\delta z) \propto |\delta z|^2$
when $|\delta z| \rightarrow 0$ \cite{fyo99,yan}. 
Thus one concludes 
that $O(z_1,z_2)$ must converge
to a constant as $|\delta z|$ approaches zero.
Since the crossover from Eq. (\ref{eq:resultO}) to 
constant behaviour occurs at
$|\delta z|^2 \simeq  (\pi d_0 N)^{-1}$,
one estimates, to lowest order in $m$
\be
\label{eq:resultnp}
\frac{O(z_0,z_0+\delta z)}{ (d_0 N)^2}\simeq
\frac{m^2}{g}\qquad
\mbox{for $|\delta z|^2 \ll (\pi d_0 N)^{-1}$}\,.
\ee
Eqs. (\ref{eq:resultO}) and (\ref{eq:resultnp}) 
are consistent with the assumption that
eigenvector correlations in the scattering
ensemble (\ref{eq:defJ}) are {\em universal}
in that $O(z_0,z_0+\delta z)$ is given,
after suitable rescaling and well within
the support of the density of states, by the
corresponding expression derived
in Ref. \cite{cha98a} for Ginibre's
ensemble: defining  \cite{note}
$\delta\tilde z \equiv \delta z \sqrt{\pi d_0 N}$
one may expect, to lowest order in $m$
\begin{eqnarray}
\label{eq:prop}
\frac{O(z_0,z_0+\delta z)}{(d_0 N)^2}\\
&&\hspace*{-1.5cm}\simeq -\frac{m^2}{g} \frac{1}{|\delta \tilde z|^{4}}\,
\left[1-(1+|\delta \tilde z|^2)\,\exp(-|\delta \tilde z|^2)\right]\,.
\nonumber
\end{eqnarray}
This expression interpolates between
(\ref{eq:resultO}) and (\ref{eq:resultnp}).

It must be pointed out that Eqs. (\ref{eq:resultnp})
and (\ref{eq:prop})
cannot be valid for very small values of $m=O(1/N)$,
where the ensemble (\ref{eq:defJ}) deviates very little from
the classical Gaussian unitary ensemble of random
Hermitian matrices \cite{meh67}. Spectral correlations
in this situation have been analyzed in detail
in Refs. \cite{fyo99,fyo97} where it was shown
how the crossover from non-Hermitian
to Hermitian ensembles may be characterized.

\begin{figure}
 \mbox{}\\[1cm]
 \centerline{\includegraphics[clip,width=0.4\textwidth]{o12.eps}}
 \mbox{}\\
 \caption{\label{fig:o12} Shows
 ${\sf Re}\,O(z_0,z_0+\delta x)/(d_0N)^2$ as a function
 of $\delta\protect\widetilde x = \delta x \,\sqrt{\pi d_0 N}$;
 for $m=0.1$,$\gamma = 0.5$, $N=1600$ ($\triangle$),
 $N=800$ 
 ($\hbox{\protect\psfig{file=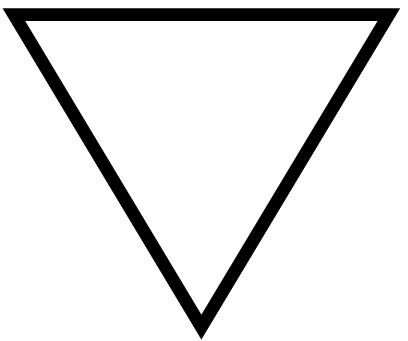,width=0.275cm}}$), 
 $N=400$ ($\circ$), $N=200$ ($\Box$) and $N=100$
 ($\Diamond$).
 Also shown is the analytical estimate according to
 Eq. (\ref{eq:resultO}) (dashed line), and Eq. (\ref{eq:prop})
 (solid line).}
 \end{figure}

{\em Numerical results}. 
We have verified the validity of Eq. (\ref{eq:prop}) 
using numerical simulations of 
of the ensemble (\ref{eq:defJ}),
for $m=0.1$,$\gamma=0.5$ and
$N=100,200,400,800$ and $1600$.
Fig.~\ref{fig:o12} shows
$O(z_0,z_0+\delta z)/(d_0 N)^2$
as a function of $\delta\tilde z$.
We observe that
the numerical results
converge to Eqs. (\ref{eq:prop},\ref{eq:resultO}).
Convergence with increasing values of $N$
is much faster for small values of $|\delta \tilde z|$
than for large values of $|\delta \tilde z|$.
In fig.~\ref{fig:o12}, the
scale of the $x$-axis differs
from that of fig.~\ref{fig:comp}(a) and differences
between (\ref{eq:resultO}) and the full
result -- as obtained within the self-consistent
Born approximation --
are not visible here.

We have also performed simulations for the modified
ensemble (\ref{eq:defJ2}). The results are very similar
to those displayed in Fig. \ref{fig:o12} (not shown).

{\em Conclusions.} In this paper
we have calculated the eigenvector
correlator $O(z_1,z_2)$ for the
ensemble (\ref{eq:defJ}) using
the self-consistent Born approximation,
and for both ensembles (\ref{eq:defJ})
and (\ref{eq:defJ2}) using numerical
simulations. Our results imply
that  eigenvector correlations
in these ensembles
are locally given by a universal law  --
after suitable rescaling of the complex
energies.  One may thus expect that
local eigenvector correlations in more general
ensembles (such as ensembles of random
Fokker-Planck operators \cite{cha97}) may be described
by the law derived in \cite{cha98a}.
It has been pointed out that such correlations may determine
transient features in the dynamics
of such systems \cite{meh99b}. 
The results found here may be of direct relevance
for quantum scattering systems \cite{lasers,rotter}.

\end{multicols}

\begin{thebibliography}{10}
\bibitem{note} This rescaling differs from that
 used for the spectral
 correlation function in Ref. \cite{fyo99} by a factor
of $\sqrt{N}$ because in \cite{fyo99}
the case $N\rightarrow \infty$ with fixed
$M$ was considered.
\end{thebibliography}
\end{document}